\documentclass[twocolumn, twocolappendix]{aastex631}
\usepackage{newtxtext,newtxmath}
\usepackage{enumerate}
\usepackage{bm}
\usepackage{xcolor}
\usepackage{graphicx}	
\usepackage{amsmath}	
\usepackage{float}
\usepackage{soul}
\setstcolor{red}
\usepackage{color}
\newcommand{\add}[1]{#1}

\newcommand{\KIAA}{\affiliation{Kavli Institute for Astronomy and
Astrophysics, Peking University, Beijing 100871, China}}
\newcommand{\DOA}{\affiliation{Department of Astronomy, School of Physics,
Peking University, Beijing 100871, China}}

\newcommand{\NAOC}{\affiliation{National Astronomical Observatories,
Chinese Academy of Sciences, Beijing 100012, China}}

\received{June 1, 2022}
\revised{June 1, 2022}
\accepted{June 1, 2022}

\submitjournal{ApJ}
\shorttitle{Realistic Detection and Early Warning of BNSs}
\shortauthors{C. Liu et al.}

\graphicspath{{./}{figures/}}

\begin{document}

\title{Realistic Detection and
Early Warning of Binary Neutron Stars with Decihertz Gravitational-wave
Observatories}

\correspondingauthor{Lijing Shao}
\email{lshao@pku.edu.cn}
\author[0000-0001-7649-6792]{Chang Liu}\DOA\KIAA
\author[0000-0001-7402-4927]{Yacheng Kang}\DOA\KIAA
\author[0000-0002-1334-8853]{Lijing Shao}\KIAA\NAOC

\begin{abstract}

We investigated the detection \add{rates} and \add{early warning parameters} of binary neutron star (BNS)
populations with decihertz gravitational-wave observatories in a realistic
detecting strategy.
Assuming 4 years' operation of B-DECIGO, we \add{based on parameter precision to classify the detectable BNSs} into three categories: (a) sources that merge within 1~year, which could
be localized with an uncertainty of $\Delta\Omega \sim 10^{0}$\,deg$^2$;  (b)
sources that merge in 1--4 years, which take up three quarters of the total events and
yield the most precise angular resolution with $\Delta \Omega\sim
10^{-2}$\,deg$^2$ and time-of-merger accuracy with $\Delta t_c\sim 10^{-1}$\,s;
and (c) sources that do not merge during the 4-yr mission window, which enable possible
early warnings, with $\Delta \Omega\sim 10^{-1}$\,deg$^2$ and $\Delta t_c\sim
10^{0}$\,s.  Furthermore, we compared the pros and cons of B-DECIGO
with the \add{third-generation ground-based detectors}, and explored the prospects of detections using 3
other decihertz observatories and 4 BNS population models. In realistic
observing scenarios, we found that decihertz detectors could even provide
early-warning alerts to a source decades before its merger while their
localizations are still \add{as accurate as ground-based facilities}. Finally we
found a decrease of events when considering the confusion noise, but
this could be partially solved by a proper noise subtraction.

\end{abstract}


\section{Introduction}
\label{sec:intro}

The detection of gravitational waves (GWs) from the binary neutron star (BNS) inspiral GW170817 
has opened an exciting era of multi-messenger
astronomy \citep{LIGOScientific:2017vwq, LIGOScientific:2017ync}. In combination
with the following electromagnetic (EM) counterparts, joint detection of BNS
mergers by GW and EM facilities provided us unprecedented
opportunities to explore many fundamental questions
\citep{Sathyaprakash:2019yqt}, including dense matter properties in extreme
conditions with associated high-energy astrophysical processes
\citep{LIGOScientific:2017pwl, LIGOScientific:2018cki}, tests of general
relativity in the strong-field regime \citep{LIGOScientific:2017zic,
LIGOScientific:2018dkp}, and the cosmic expansion with
the ``standard siren'' technique \citep{Schutz:1986gp, LIGOScientific:2017adf}.

The prerequisite for a successful multi-messenger detection is the prompt
communication of source location and other properties from GW data to EM
telescopes.  For GW170817, the short $\gamma$-ray burst (GRB),
GRB\,170817A, was observed $\sim 1.7$\,s after the BNS merger, while the
localization from LIGO/Virgo took hours \citep{LIGOScientific:2017zic}.  People
have been studying and improving the low-latency GW triggers
\citep{Hooper:2011rb, Nitz:2018rgo, LIGOScientific:2019gag}. They have also
assessed the detection and early warning abilities for compact binary mergers
observed by current and future ground-based GW detectors, including LIGO, Virgo,
and KAGRA \citep{KAGRA:2013rdx}, as well as the third-generation \add{(3G)} detectors, the
Einstein Telescope (ET) and Cosmic Explorer \add{\citep[CE; see e.g.,
][]{Chan:2018csa, Sachdev:2020lfd, Nitz:2021pbr, Singh:2021zah,
Magee:2022kkc,Borhanian:2022czq}}.

While the real-time detections and early warnings from ground-based GW detectors
have been closely examined, less attention has been paid to decihertz GW
detectors. \add{Current decihertz detector proposals \citep{Izumi:2021vrp} include the space-based interferometers \citep{TianGo, Sedda:2019uro, Yagi:2011wg}, the atomic inteferometers \citep{Graham:2017pmn, Zhao:2021bjw}, and the lunar GW detectors \citep{Jani:2020gnz}. With space-based decihertz detectors, the research on detections rates \citep{Piorkowska-Kurpas:2020rfy, Seto:2001qf, Geng:2020wns, Cao:2021jpx} and parameter estimation \citep{Nair:2018bxj, Isoyama:2018rjb,
Nakano:2021bbw} uses simplified detection strategies. Few studies focused on the realistic detection of BNS population and the early warnings from decihertz GW detectors \citep{Liu:2021dcr}. }

Decihertz space-based GW detectors have a distinct advantage over ground-based
ones, for which the GW signals from BNS systems could stay days to
years in the decihertz band \citep{Isoyama:2018rjb}. 
By virtue of this,
decihertz detectors can observe sources that are not only about to merge, but
also solely in the inspiral stage for the duration of observation. For such sources, decihertz detectors could
gather enough information from their pre-merger stages and provide the locations
and times of merger in advance to EM facilities. \add{Even} after the
detectors end their missions, their legacies on source information remain
valuable for early warnings for a couple of years. 

\add{Another motivation to study the early warnings from decihertz detectors is that BNS systems detected by decihertz detectors will very likely be multiband sources \citep{Mandel:2017pzd}, as they will eventually merge in the kilohertz band within a reasonable time. Multiband detections would strongly improve the parameter precision, test gravity theories, and boost the success of multi-messenger astronomy \citep{Sesana:2016ljz, Vitale:2016rfr, Grimm:2020ivq, Gerosa:2019dbe, Liu:2020nwz, Jani:2019ffg, Klein:2022rbf}.}

What a pre-merger alert needs most is the information of source location and time of
merger. \add{Theoretically, the localization depends on
the detector's trajectory baseline and the orientation of the source \citep{Liu:2021dcr}. The time of merger accuracy scales inversely with the detector's frequency bandwidth and the source's signal-to-noise ratio \citep[SNR;][]{Grover:2013sha}.}
However, different signal durations, population distributions,  detector designs, and mission time make this problem complex. 

The aim of this paper is to extend early
studies on decihertz BNS early warnings by simulating detections in real
observing scenarios, and exploring the distributions of the accuracy in
localization and timing. \add{We propose a strategy of realistic detections and investigate early warning properties of 4 BNS population models with 4 space-based decihertz GW observatories using the Fisher information matrix \citep{Finn:1992wt}. Under the combined effects of mission time and signal duration, we proposed a new classification of BNS sources by dividing them into 3 categories according to the merger time and parameter precision.} In addition, we compare the detection performances with the 3G detectors, and discuss the influence of the confusion noise. 


The paper is organized as follows.
In Sec.~\ref{sec:setups}, we introduce the population models, and relavent GW detectors.
\add{In Sec.~\ref{sec:abc} we propose the realistic detecting strategy and provide a new classification of BNS sources.} In Sec.~\ref{sec:BDEC}, taking B-DECIGO
\citep{Kawamura:2020pcg} and one population model as an
example, \add{we illustrate typical characteristics of the three BNS categories with distinctive localization and timing abilities, then briefly discuss the implication on EM detections.}
Section~\ref{sec:ET} compares the predictions from B-DECIGO with that from \add{3G detectors},
and Secs.~\ref{sec:dets} and~\ref{sec:pops} compare the detection prospects
using other decihertz detectors and other population models, respectively.  In
Sec.~\ref{sec:conf}, we briefly discuss the influence of the GW foreground from
compact binaries on detection rates. Section~\ref{sec:sum} concludes the study. 
Appendices give more information on the merger rate calculation and GW foreground from unresolved double compact objects.
Throughout this paper, we use geometrized units in which $G=c=1$. 

\section{\add{BNS Populations and GW Detectors}}
\label{sec:setups}

In this section, we will introduce the BNS population models (Sec.~\ref{subsec:pop})
and decihertz detectors  (Sec.~\ref{subsec:det}) that we adopted in this work.

\subsection{BNS population models} \label{subsec:pop}

\begin{figure}
	\includegraphics[width=9cm]{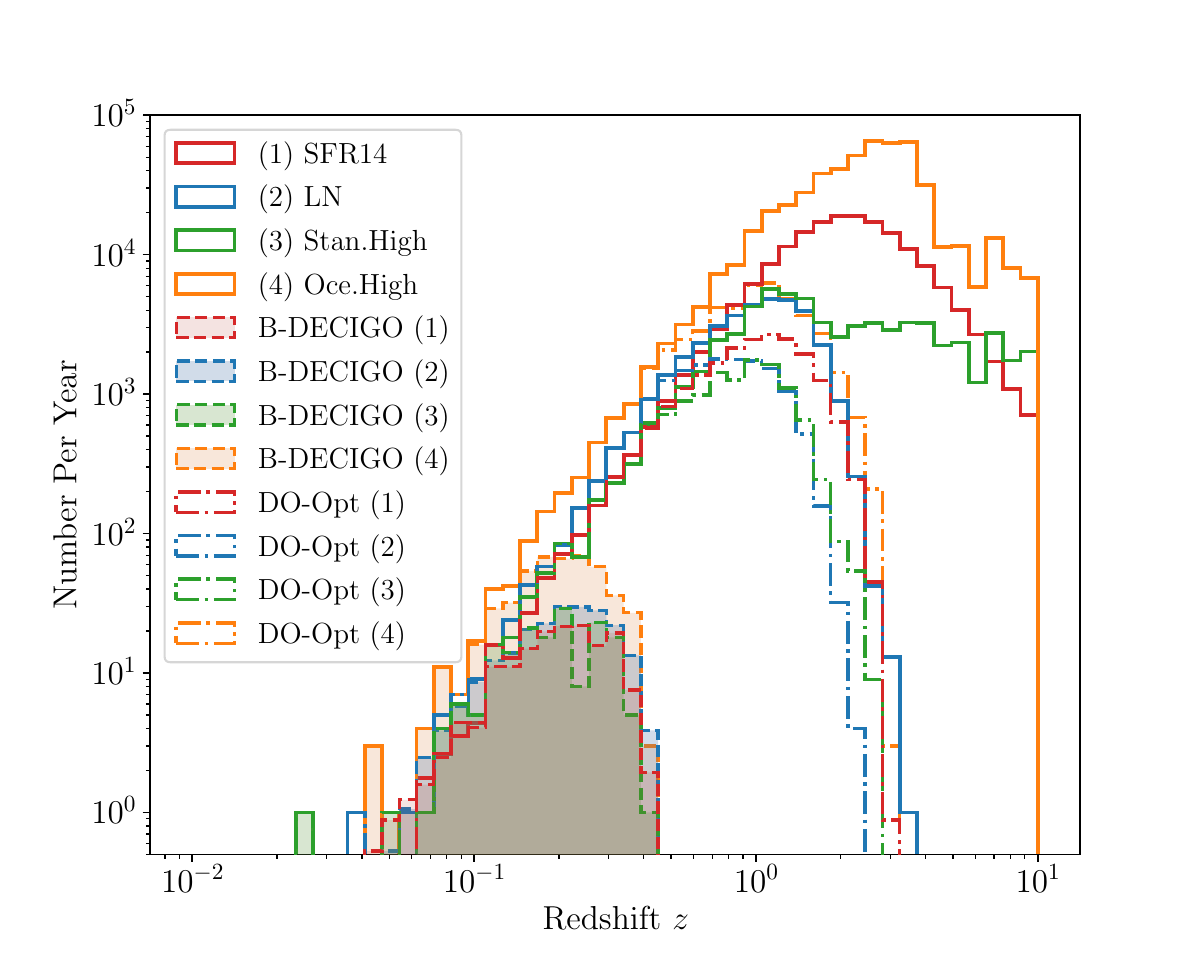}
	\caption{Solid histograms show the total number of BNSs per year from population models. 
	\add{Dash-shaded} and dash-dotted histograms show sources that will merger in the fourth year of a 4-yr mission,  to be detected by B-DECIGO and DO-Optimal,
	  respectively.}
    \label{fig:pops}
\end{figure}

For the cosmic evolution of BNS merger rate, there are various types of
population models \citep[for a review, see ][]{Mandel:2021smh}. In our work, we
have implemented 4 representative classes of them. Their distributions are given
in Fig.~\ref{fig:pops} with solid histograms, and the models---abbreviated as ``SFR14'', ``LN'', ``Stan.High'', and
``Oce.High''---are introduced as follows:
\begin{enumerate}[(1)]	
	\item {\bf SFR14}---We assumed that the merger rate evolves with redshift
	following the fitting formula of star formation rate (SFR) from
	\citet{Madau:2014bja}, and we adopted a local merger rate
	$\mathcal R_0=44$\,Gpc$^{-3}$\,yr$^{-1}$ based on the ``PDB (ind)'' model which is informed by
	events in GWTC-3 of compact binary mergers \citep{LIGOScientific:2021psn}.
	\item {\bf LN}---Many population models are constructed based on a delay
	time superposed on the SFR, such as the Gaussian delay model, power-law
	delay model \citep{Virgili:2009ca}, and log-normal delay model
	\citep{Wanderman:2014eza}. According to the observations of short GRBs,
	log-normal delay model is favored \citep{Sun:2015bda}. We adopted the
	dimensionless redshift distribution of the log-normal delay model in Eq.~(A8) of
	\citet{Zhu:2020ffa}, and a local merger rate $\mathcal R_0=44$\,Gpc$^{-3}$\,yr$^{-1}$.
	\item {\bf Stan.High}---We adopted the  ``Stan.High'' and ``Oce.High'' models
	from
	\citet{Dominik:2013tma}.\footnote{\url{https://www.syntheticuniverse.org}}
	These models have been applied in many other studies \citep{Dominik:2014yma,
	Ding:2015uha,Piorkowska-Kurpas:2020rfy}. ``Stan.High'' is the standard model
	with a ``high-end'' metallicity evolution.
	\item {\bf Oce.High}---``Oce.High'' is the ``Optimistic common envelope''
	model with a ``high-end'' metallicity evolution \citep{Dominik:2013tma}. 
\end{enumerate}

Although different population models evolve differently with redshift, most of
the population synthetic codes start with the SFR of \citet{Madau:2014bja} and
some of them have similar evolving trends \citep{Baibhav:2019gxm}.  Therefore,
similarly to \citet{Nitz:2021pbr}, we choose SFR14 as our fiducial model.
Without specific mention, our calculations are based on it. Our redshift cutoff
is at $z=10$ and we transform redshift to luminosity distance $D_L$ based on
the $\rm \Lambda CDM$ model with the matter density parameter $\Omega_M =
0.315$, the dark-energy density parameter $\Omega_\Lambda = 0.685$, and the
Hubble constant $H_0 = 67.4\, \rm km\, s^{-1}\, Mpc^{-1}$
\citep{Planck:2018vyg}.

We focus on a typical BNS system where the two component masses, $M_1$ and
$M_2$, are both $1.4 \,M_\odot$. According to \citet{Nitz:2021pbr}, the
localization results can be  rescaled to apply to sources with other masses or
local merger rates.  For other parameters, we choose the dimensionless tidal
deformability $\tilde \Lambda=675$, the spins of  neutron stars $\chi_{1,2}=0$,
the source direction angles, $\cos\bar\theta_S \in \mathcal{U}(-1, 1)$ and
$\bar\phi_S\in \mathcal{U}(0,2\pi)$, and the angular momentum direction angles,
$\cos\bar\theta_L\in \mathcal{U}(-1, 1)$ and $\bar\phi_L\in
\mathcal{U}(0,2\pi)$, where $\mathcal{U}(\cdot, \cdot)$ denotes a uniform
distribution.

\subsection{\add{GW detectors}}  \label{subsec:det}

\begin{figure*}
\centering
	\includegraphics[width=13cm]{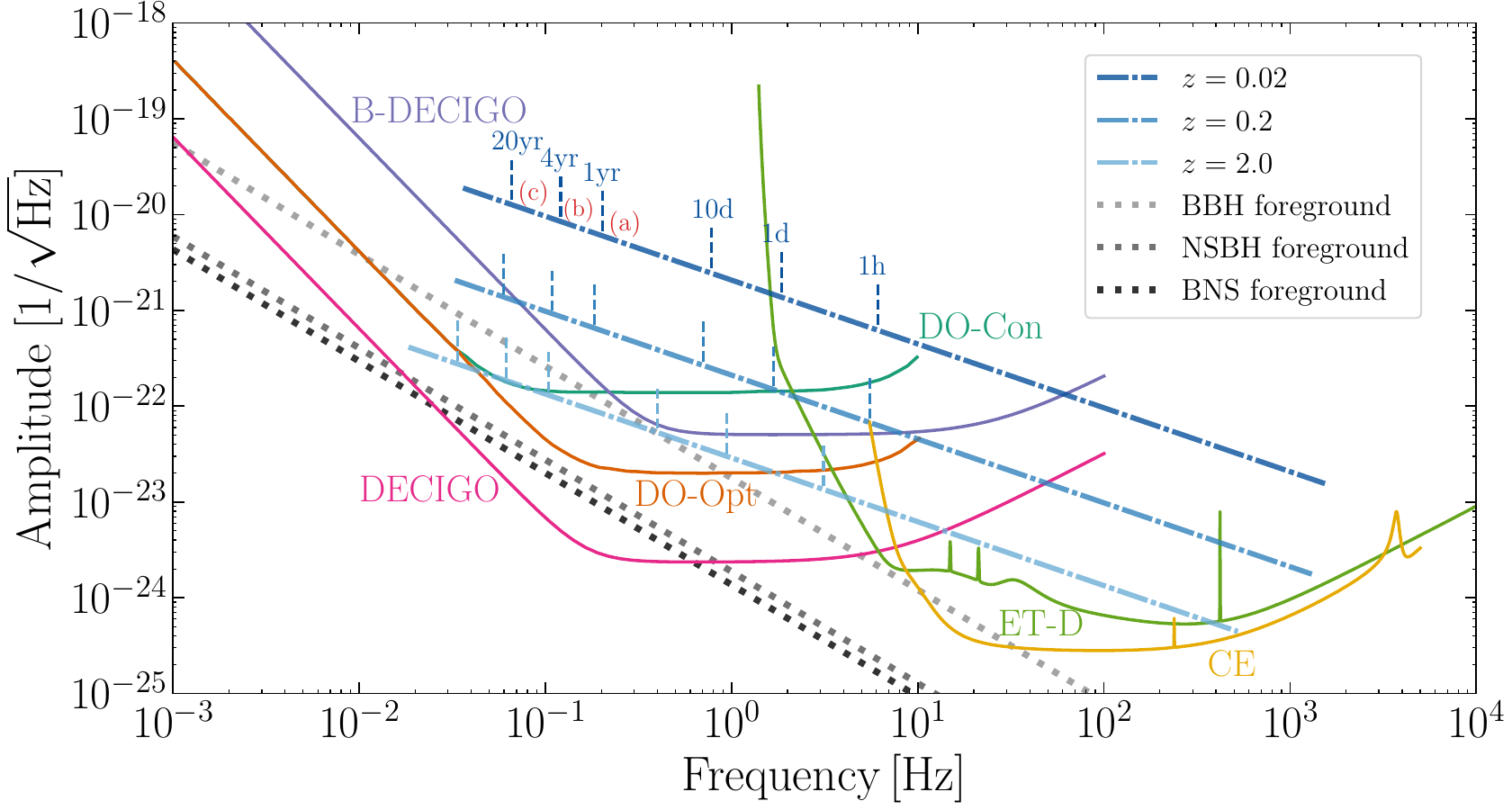}
	\caption{\add{The amplitude of the sources and detector noises.} The
	sky-averaged effective noise $\sqrt{{S_{\rm n}^{\rm eff}}(f)}$ of various
	detectors [see Eq. (29) of \citet{Liu:2020nwz}] is given in solid lines; the
	pre-subtracted BNS/NSBH/BBH foregrounds from model ``Stan.High'' are given in dashed grey lines; the
	simulated BNS signals, $2\sqrt{f}|\tilde{h}_+(f)|$, at redshift $z=$ 0.02,
	0.2, and 2.0 are illustrated with dash-dotted blue lines. Source signals are plotted with a duration of 100 years. For
	each source, the short dashed vertical lines mark the times before
	coalescence. Categories (a), (b), and (c) are annotated on the illustrated source at
	$z=0.02$.}
	\label{fig:curve}
\end{figure*}

We use 4 space-based decihertz GW detectors to explore their performance: (i) the baseline mission B-DECIGO
\citep{Isoyama:2018rjb,Kawamura:2020pcg}, (ii) the full-scale mission DECIGO
\citep{Yagi:2011wg, Kawamura:2011zz}, (iii) a conservative example
DO-Conservative, and (iv) an optimal example DO-Optimal \citep{Sedda:2019uro,
Sedda:2021yhn}. \add{Later in Sec.~\ref{sec:ET}, to compare with the next-generation ground-based detectors, we calculate the detection prospects of the B-DECIGO sources to be observed by ET alone as well as a combination of ET and CE (``ET+CE''). We assume that ET \citep{Hild:2010id} consists of 3 detectors located in the same place as the Virgo detector in Italy, and CE \citep{Evans:2016mbw} consists of 2 detectors located at Livingston and Hanford sites as LIGO detectors, respectively. The noise curves of the aforementioned detectors are shown in Fig.~\ref{fig:curve}. Readers are referred to Sec.~2.2 of \citet{Liu:2021dcr} for details. }

\section{\add{BNS EARLY-WARNING CATEGORIES}}
\label{sec:abc}

Assuming BNS systems will merge following the prediction of a population model,
every year there will be approximately the same amount of merger signals.
Contrary to ground-based hectohertz
GW detectors, where signal duration $\ll$ mission lifetime, in decihertz detectors, however, BNS signals could exist from the start of the mission to the merger or even to the end of the mission. 
Except for those BNSs that merge within the first few days of the
mission, all the signals will exist in decihertz detectors long enough to
guarantee a relatively stable parameter estimation precision, especially for
localization. \add{For long-lasting
signals, the detection rates and the localization precision make no difference whether we end the
observations a few hours or a day before the merger \citep[see e.g., Fig.~7 in][]{Liu:2021dcr}.}  Therefore for
decihertz space-based detectors, ``early warning'' is time-independent to a
certain extent.  They can always provide accurate localization and 
time of merger to both GW and EM detectors for follow-ups. Therefore we pay more attention to mock the realistic detections.

\add{Our realistic detection strategy is performed as follows.} From the start of the
mission to 20 years after it, we generate BNS mergers weekly according to the population
model and then calculate their SNRs.  If the SNR > 8, we claim the detection and calculate their parameter precisions by the Fisher matrix method. \add{The integration interval for SNR and Fisher calculation depends on BNS categories and we will discuss it in the next paragraph.} We follow the setups in \citet{Liu:2021dcr} with slight revisions: parameters
used in the Fisher information matrix are $\bm{\Xi} = \big\{  {\cal M}, \eta, t_c, \phi_c,
D_{L}, \bar \theta_S, \bar\phi_S,\bar\theta_L, \bar\phi_L \big\} \,$, where
${\cal M}$ is the chirp mass; $\eta$ is the symmetric mass ratio; $\phi_c$ and
$t_c$ are the phase and time at the coalescence.  Our attention focuses on the
estimation of the accuracy of angular resolution, $\Delta \Omega$, and time of merger, $\Delta t_c$.  

We conservatively set the mission time of the decihertz space-based detectors as $T_{\rm Mission} = 4$ years and denote the BNS merger time from the start of the mission as $t_{c_0}$. Based on the distribution of $\Delta \Omega$ and $\Delta t_c$ in our results, we classify the BNS sources into 3 categories: 
\begin{enumerate}[(a)]
\item  BNSs  that merge within 1 year ($t_{c_0} \leq 1\,$yr); 
\item  BNSs that merge in 1 to  4 years ($1\,{\rm yr} < t_{c_0} \leq 4\,$yr);
\item  BNSs that only inspiral within the whole 4-yr observational span
($t_{c_0} > 4\,$yr).
\end{enumerate}
We use ``(a)'', ``(b)'', and ``(c)'' to denote these categories, and they are
annotated in Fig.~\ref{fig:curve} for a BNS signal at $z=0.02$. 
\add{During
the calculation of the SNR and the Fisher matrix, we need to calculate the inner
product, which is an integration in the frequency domain, from ${f_{\rm in}}$ to
${f_{\rm out}}$.
Catogories (a) and (b) contain sources that will merge within $T_{\rm Mission}$ ($t_{c_0} \leq T_{\rm Mission}$). Their signal durations depend on  $t_{c_0}$. The larger the $t_{c_0}$, the more time they will stay in the detector. Thus, they enter the detector at ${f_{\rm in}}=\left( t_{c_0}/{5}\right)^{-3/8}{\cal M}^{-5/8} / 8 \pi$ and leave the detector at the frequency upper limit of the detector, $f_{\rm out}=f_{\rm high}$.
Category (c) contains sources that only inspiral throughout $T_{\rm Mission}$ ($t_{c_0} > T_{\rm Mission}$). They enter the detector at ${f_{\rm in}}=\left( t_{c_0}/{5}\right)^{-3/8}{\cal M}^{-5/8} / 8 \pi$ and leave the detector 4 years later when the detector ends its mission, with ${f_{\rm out}}=\big[ (t_{c_0} - 4\,{\rm yr}) /{5}\big]^{-3/8}{\cal M}^{-5/8} / 8 \pi$.} For category (c) their signals could be analyzed after the mission, which means detectors could even provide ``legacy'' information after they stop operating.

Note that we generate BNS sources weekly so that a shorter sampling period
would cause large fluctuations in event numbers, and a longer sampling period
would lead to a biased parameter estimation result.  Moreover, the event numbers may be decimals due to normalization.
Following the evolution of each population model, we simulate plenty of
sources by adjusting $\mathcal R_0$ in Eq.~\eqref{eq:rate}, and when presenting
the results we normalize to the model's local merger rate,  e.g., ${\cal R}_0 = 44$\,Gpc$^{-3}$
yr$^{-1}$ for the ``SFR14'' model. It is referred as ``normalized simulation'' in
later sections.

\section{CHARACTERISTIC OF BNS EARLY-WARNING CATEGORIES}
\label{sec:BDEC}

\begin{figure}
	\includegraphics[width=8.8cm]{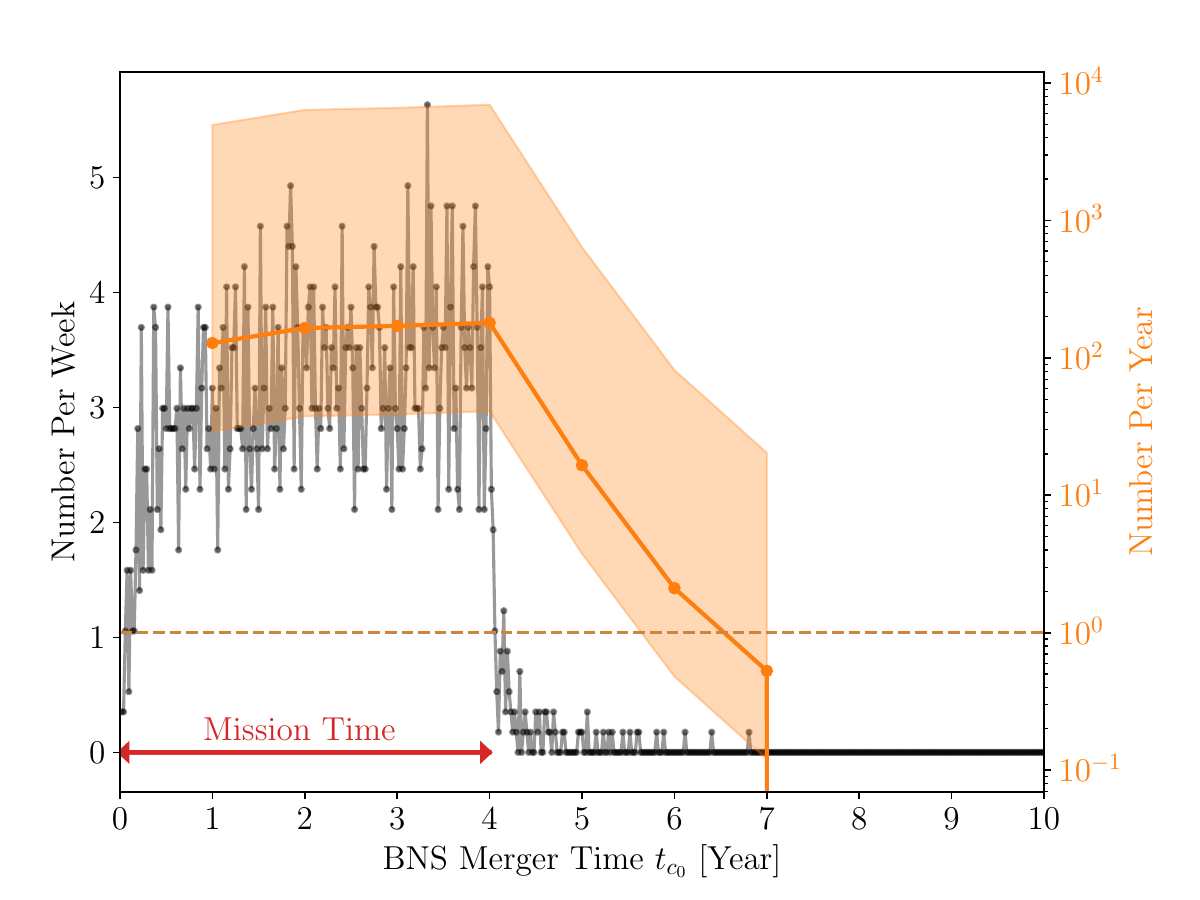}
	\caption{Black symbols show the detection number per week for B-DECIGO as a function of BNS
	merger time $t_{c_0}$ since the observation starts. We assume that the
	mission lasts for 4 years, and have used the population model ``SFR14''. The orange dots and shaded area,
	with labels on the right hand side, indicate respectively the yearly detection number and the range of BNS local merger
	rate inferred from GWTC-3, namely  in the range of $[10, 1700] \,{\rm
	Gpc}^{-3}\,{\rm yr}^{-1}$ \citep{LIGOScientific:2021psn}. The dashed horizontal brown
	line gives a yearly detection of only 1 BNS. Sources that cannot merge
	within 7 years after the launch of B-DECIGO could not be detected.}
    \label{fig:BDEC_number}
\end{figure}
\begin{figure}
	\includegraphics[width=8.5cm]{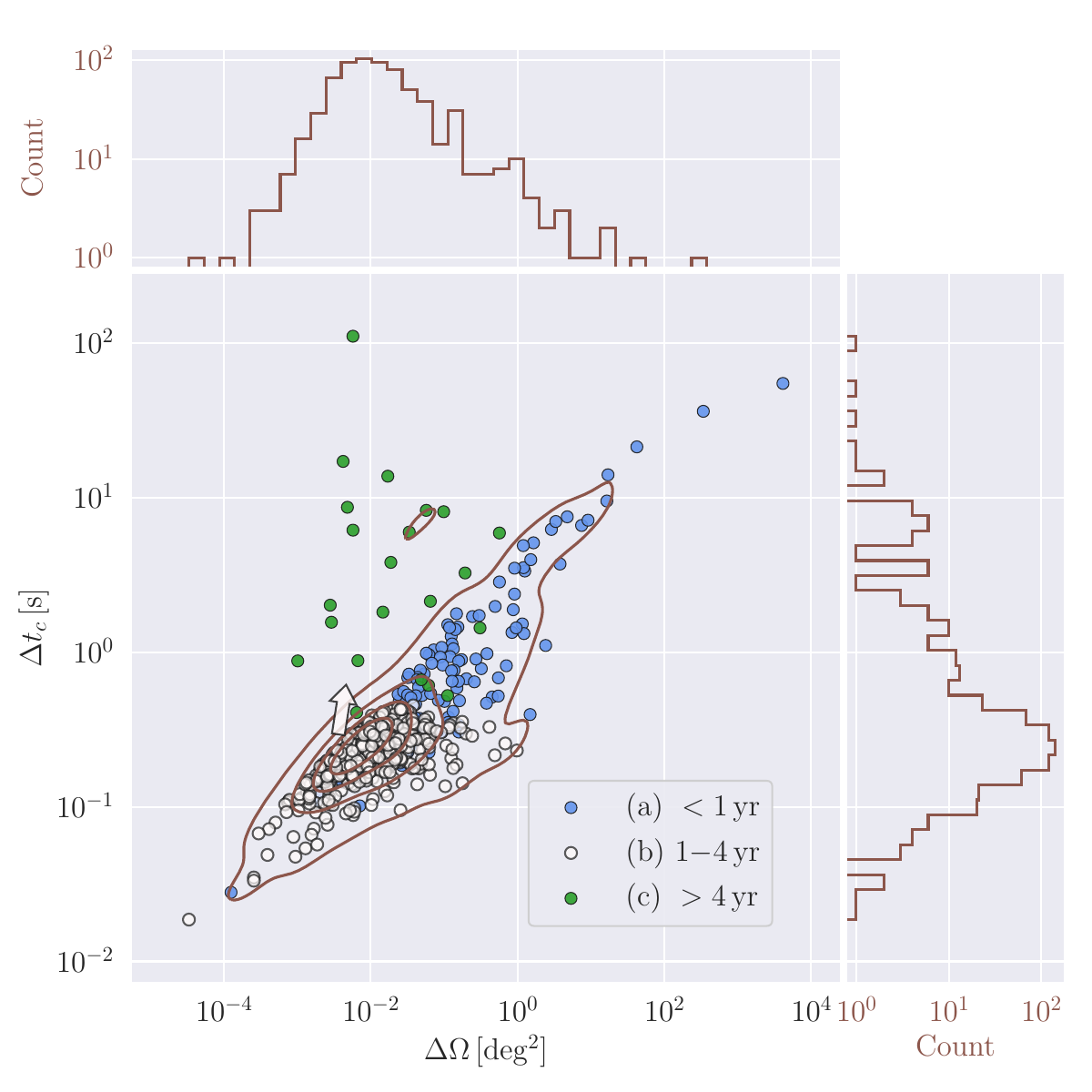}
	\caption{The $\Delta\Omega$-$\Delta t_c$ distribution of the sources that
	could be detected during the operation of B-DECIGO. Population model
	``SFR14'' is adopted. The categories (a), (b), and (c) are plotted in blue,
	white, and green circles, respectively. The white arrow shows the direction
	in which the parameter precision of category (b) moves, if early warning is
	required. The 4 brown contour lines delimit 5 iso-proportions of the number
	density.}
    \label{fig:BDEC_dist}
\end{figure}
\begin{figure}
	\includegraphics[width=8.8cm]{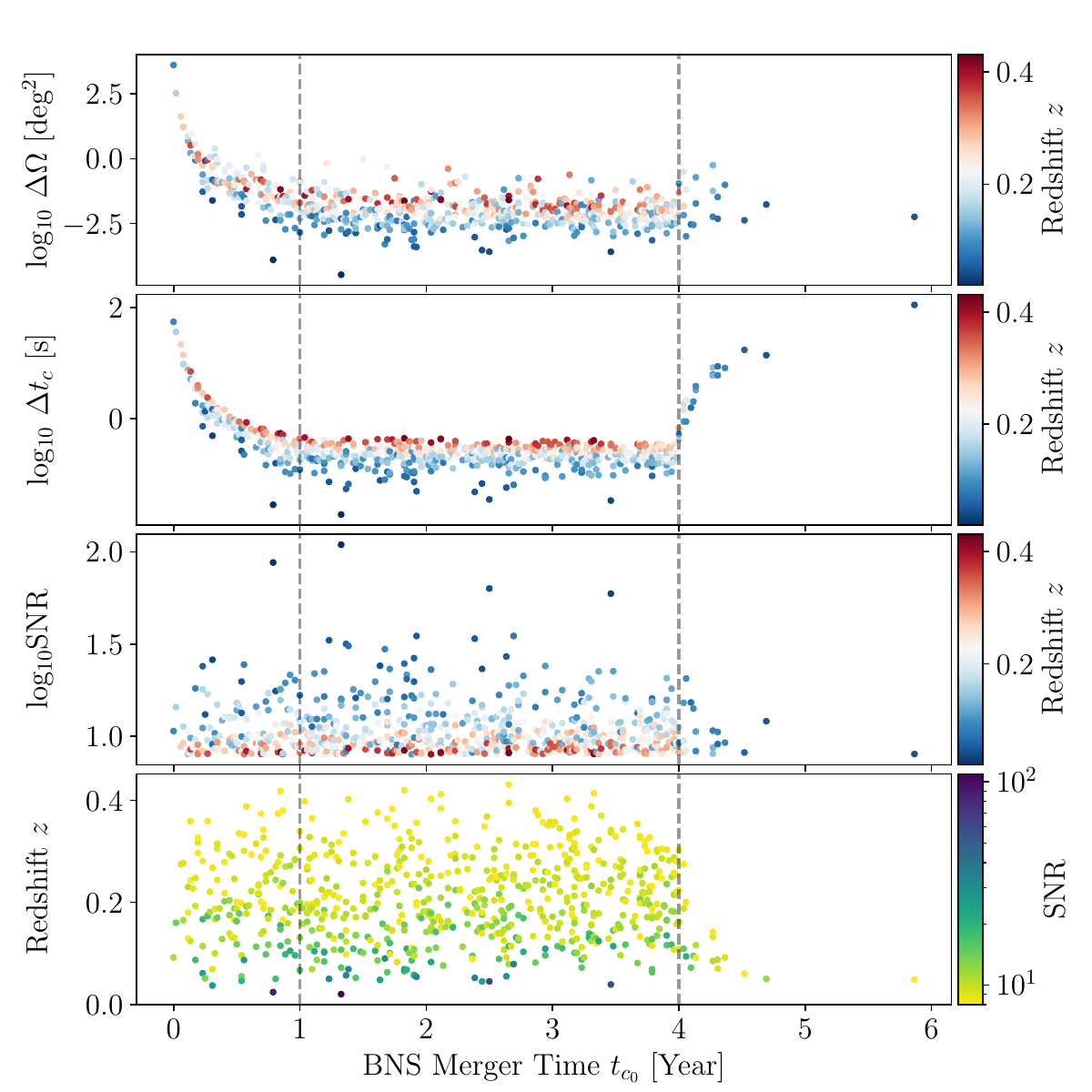}
	\caption{The angular resolution ({\it top}), time of merger accuracy ({\it
	second panel}), SNR ({\it third panel}), and redshift ({\it bottom}) of the
	sources detected by B-DECIGO as a function of $t_{c_0}$.  The dots are
	plotted weekly since the calculation is performed with such a cadence.
	Colors indicate the redshift or SNR of these sources. Two vertical dashed
	lines divide BNSs into, from left to right, categories (a), (b), and (c).}
    \label{fig:BDEC_4panel}
\end{figure}

\begin{table*}
\def\arraystretch{1.25}
	\centering
	\caption{Event numbers detected by 4 decihertz GW detectors for different
	observation periods since the start of the observation. The percentage in the brackets shows the ratio of it to the total number of detections in 20 yrs.
	We have used the ``SFR14'' population model. }
	\label{tab:Number/yr}
	\setlength{\tabcolsep}{1.15mm}{\begin{tabular}{lcccccccc} 
		\hline
		\hline
		  Detector & 
		  $z_{\rm max}$ &
		  $t_{c_0}\in(0,1]\,$yr & 
		  $t_{c_0}\in(1,2]\,$yr &
		  $t_{c_0}\in(2,3]\,$yr & 
		  $t_{c_0}\in(3,4]\,$yr & 
		  $t_{c_0}\in(4,5]\,$yr &
		  $t_{c_0}\in(5,10]\,$yr &
		  $t_{c_0}\in(10,20]\,$yr\\
		 \hline
B-DECIGO & $ \sim 0.45$ & 128 (19\%) & 165 (25\%)& 171 (26\%)  & 180 (27\%) & 16.5 (2.5\%) & 2.64 (0.4\%) & 0 (0\%) \\
DO-Conservative & $\sim 0.26$ & 10.4 (3.7\%) & 25.8 (9.3\%) & 36.8 (13\%)  & 50.5 (18\%) & 32.0 (11\%) & 70.5 (25\%)& 53.0 (19\%)  \\
DO-Optimal & $\sim 3.0$ & 9610 (13\%) & 17200 (23\%) & 19400 (25\%) & 20400 (27\%)& 5620 (7.4\%) & 3230 (4.2\%)&  535 (0.7\%) \\
DECIGO & $>10$ & 175000 (15\%)& 175000 (16\%)& 175000 (16\%)&  175000 (16\%)& 140000 (12\%)& 223000 (20\%) &  66700 (5.9\%) \\
		\hline
	\end{tabular}}
\end{table*}

In this section, taking the detector B-DECIGO and population model ``SFR14'' as
an example, we present different characteristics with categories (a), (b), and
(c) defined in Sec.~\ref{sec:abc}. Our main conclusion is that, the statistical properties of accuracy in localization and
time of merger for these 3 categories have distinct and unique features, which are determined by the synergy of sensitivity curves, source merger time, and the mission time. It is better to consider them separately when studying them.     

We first introduce
some characteristics in our figures and tables.  Figure~\ref{fig:BDEC_number} shows the weekly
and yearly detection rates of the BNSs, while Table~\ref{tab:Number/yr} records the yearly detections. Those results are based on normalized simulations.
Figures~\ref{fig:BDEC_dist} and \ref{fig:BDEC_4panel}, on the other hand, show
the distributions of the BNS sources from a single simulation, i.e., using
$\mathcal{R}_0=$ 44 Gpc$^{-3}$ yr$^{-1}$ directly.

For BNSs in category (a), their signals only stay shortly $( \leq 1\,{\rm yr})$ in
B-DECIGO. As a result, not enough information could be accumulated by the
detector to enable precise parameter estimation.  Figure~\ref{fig:BDEC_number}
shows that the weekly number of detections (SNR $> 8$) increases gradually from
less than one to about $ 4$, then becomes stable. Meanwhile, from
Figs.~\ref{fig:BDEC_dist} and \ref{fig:BDEC_4panel} we see that $\Delta t_c$ and
$\Delta \Omega$ from $t_{c_0} = 0$ to $t_{c_0} = 1\,{\rm yr}$ have an
improvement of $3$ to $4$ orders of magnitude. The reason for localization accuracy can be explained by the trajectory baseline of the
detector, which is short at the beginning of the mission, leading to an inaccurate
sky localization in category~(a).

The majority of the detectable BNS sources in B-DECIGO belong to category (b),
which yields the best and most stable parameter estimation results.  From
Table~\ref{tab:Number/yr} we notice that the yearly detection rate of this
category is more than that of category (a), and the entire category (b) takes up
78\% of the total detections. From the middle parts of
Fig.~\ref{fig:BDEC_4panel} ($1\,{\rm yr} < t_{c_0} \leq 4\,{\rm yr}$) and the white circles in Fig.~\ref{fig:BDEC_dist} we
find that the angular resolutions and timing accuracies are clustered around
$\Delta\Omega\sim 10^{-2}\,\rm{deg^2}$ and $\Delta t_c\sim 0.1\, \rm{s}$. Such a
steady distribution is unrelated to the integration time, which is the reason
why we consider all the BNSs that merge within 1 to 4 years into this category.
Note that we derive our results by integrating to $f_{\rm out}=100\,$Hz, which
is $\sim2\,$s before the actual final merger. If one wants parameter
estimation results from days earlier to execute the early warning alerts, the
white arrow on Fig.~\ref{fig:BDEC_dist} shows the direction of the changes. 
We also find that the SNR and redshift distributions are
similar for both categories (a) and (b), where sources up to $z=0.45$ can be
detected.

Category (c) contains a special kind of BNSs from
space-based detectors.  The prominent feature of category (c) is that only
nearby sources from us could be observed, judging from the last panel of Fig.~\ref{fig:BDEC_4panel} when $t_{c_0} > 4$\,yr.  Another feature is that the timing accuracy $\Delta t_c$ drops rapidly to $\sim 10^2\,$s in category~(c).
The reason would be a lack of
information due to a short integration interval in the frequency band, which is
caused by the quasi-monochromatic waves emitted by the sources that are far from
their merger stage. However, although the precision of $\Delta t_c$ in category~(c) is
relatively poor, it is still useful in most cases.
The $\Delta\Omega$ in this category is only slightly less accurate than in category (b), with a mean
value of $0.1\,\rm{deg^2}$. The baseline of B-DECIGO in the Earth orbit from a 4-yr continuous observation gives rise to such accuracy. 

For B-DECIGO, only $\sim 3\%$ of the detected sources belong to category (c),
and it cannot detect sources with $t_{c_0} \gtrsim 7\,$yr because of its
decreased sensitivity at $0.1$\,Hz. This non-detection problem does not exist in
other decihertz GW detectors such as DO-Conservative, for which $\sim 50\%$ of
the detectable sources are in category (c). We will discuss this point in
Sec.~\ref{sec:dets}. As we will see later, the typical detection features of the sources in
categories (a), (b), and (c) have the similar patterns, regardless of changes in
detectors (or sensitivity curves) and population models.  Their relative
positions on a $\Delta\Omega$-$\Delta t_c$ plot are similar to those in
Fig.~\ref{fig:BDEC_dist} and the evolution trends of the observables with source
merger time $t_{c_0}$ are similar to those in Fig.~\ref{fig:BDEC_4panel}, while
only the numbers of detections and absolute precisions alter.

\add{To distinguish binary formation channels, spins and eccentricity measurements are usually stronger indicators. The detection numbers, on the other hand, are not sufficient to determine formation channels. However, the number of detections as a function of redshift may give a hint on the time-delay distribution \citep{Baibhav:2019gxm} and a significant lack of detection also indicates certain population synthetic parameters \citep{Broekgaarden:2021efa}. We discuss the discrimination of our 4 models in Sec.~\ref{sec:pops}.}

\subsection{\add{The landscape of multiband and multi-messenger detections}}
\add{We briefly discuss the prospects of joint detections with other GW/EM telescopes regarding the 3 BNS categories. 
For category (a), their angular resolutions are not always sufficient to be covered by the field of view (FoV) of EM telescopes. Meanwhile, the signal duration is so short that a low-latency early warning is relatively difficult. Thus, of all 3 categories, the results from category (a) are  similar to those of ground-based GW detections. 
For category (b), they are the best sources for multiband and multi-messenger astronomy. Their long signal duration in decihertz detectors enables precise early warning information. The accurate parameter precisions keep the same days before the merger with angular resolutions less than typical FoVs of EM telescopes in most cases. This can hardly be achieved by ground-based GW detectors alone.
  Due to such an accurate localization and timing ability, we investigate the $\gamma$-ray burst and kilonova detection rates of sources in category (b) for various EM facilities in great detail in a separate publication \citep{Kang:2022nmz}. 
For category (c), since we obtain the source information years before their mergers with accurate localization, joint detections with both EM and ground-based GW searches would be promising. A problem might raise from the larger $\Delta t_c$, which may prevent the association with the merger signal in ground-based detectors. However, the association of other parameters, such as chirp mass and location will solve this problem.}

\begin{figure*}
\centering
	\includegraphics[width=16cm]{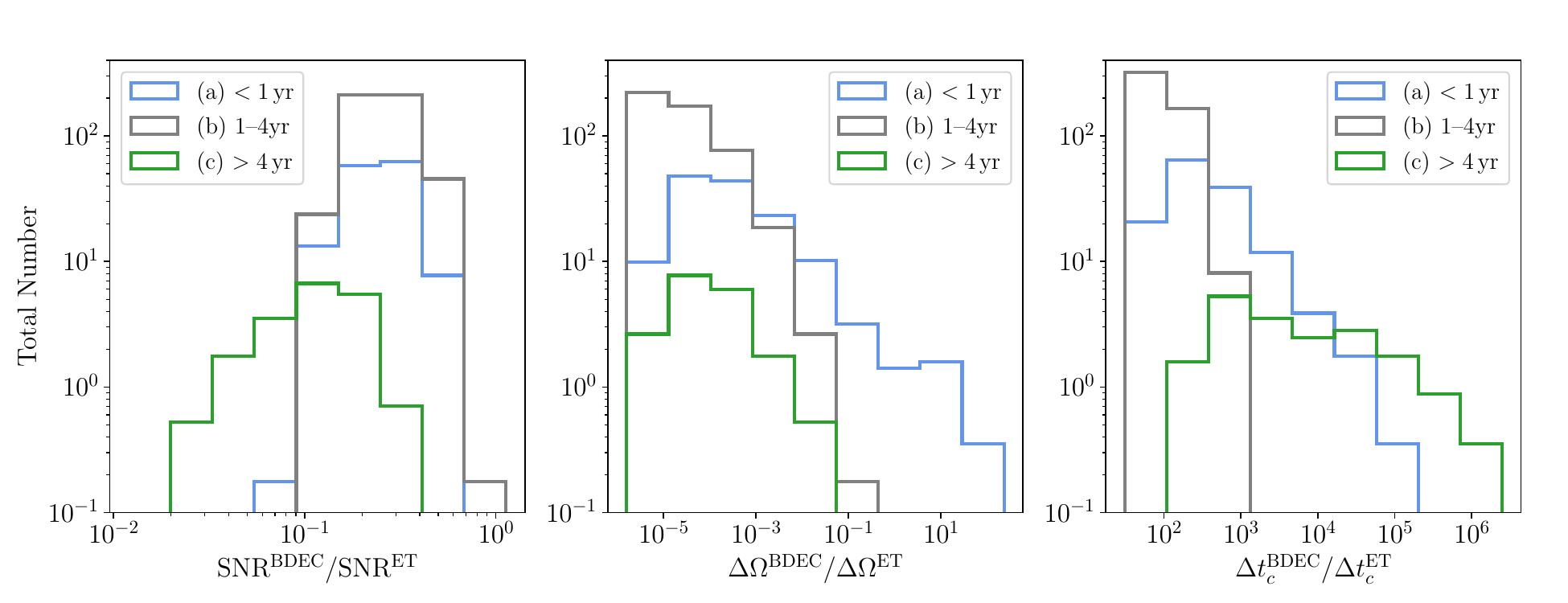}
	\includegraphics[width=16cm]{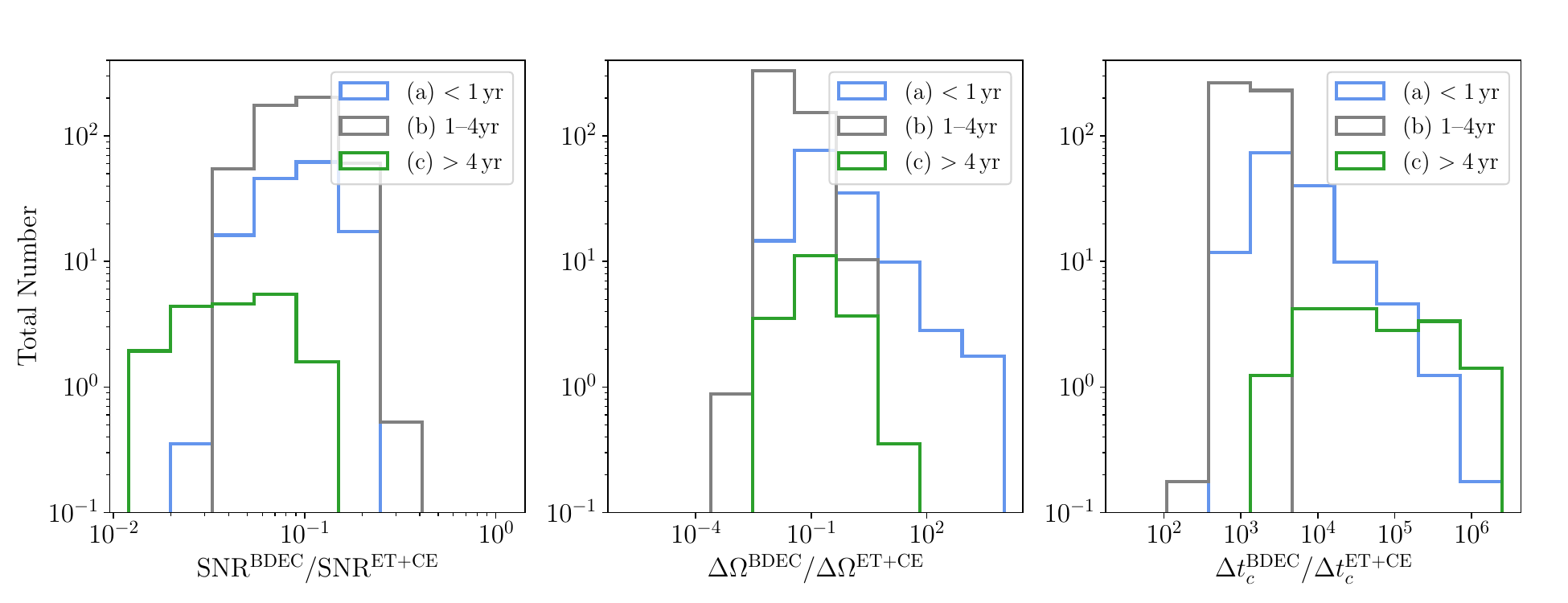}
	\caption{\add{Distributions of the ratios of SNR ({\it left}), 
	angular resolution ({\it middle}) and accuracy in the time of merger ({\it right}), for the
	sources to be detected by B-DECIGO over that by ET ({\it upper panels}) and ``ET+CE'' ({\it lower panels}). }Sources in categories (a),
	(b), and (c) are given in blue, grey, and green histograms, respectively.}
	\label{fig:ET}
\end{figure*}

\section{Comparisons with \add{the next generation ground based detectors}}
\label{sec:ET}

In this section, in order to illustrate the
pros and cons of space-based decihertz GW detectors, we compare $\Delta\Omega$, $\Delta t_c$, and SNR of the BNS
sources to be detected by B-DECIGO with that of the ET alone \add{and ``ET+CE''} \citep{Kalogera:2021bya}. 

When a BNS signal enters ground-based detectors, the merger phase becomes
prominent, such that the effective tidal deformability $\tilde
\Lambda$ and the effective spin $\chi_s$ become important to the GW
phase evolution. Therefore we also consider these parameters in the calculations
of ET \add{and CE} \citep[see e.g., ][]{Gao:2021uus, Liu:2021dcr}.

Figure~\ref{fig:ET} compares the results from B-DECIGO with that of ET \add{and ``ET+CE''}. We notice that the SNRs in ET (``ET+CE'') are a few (dozens) times higher than that of
B-DECIGO, which means that all B-DECIGO sources could be identified
by 3G detectors when they merge, regardless of their classification. This is
especially true for sources in category (c) because they are much closer to us
than sources from other categories.  In spite of this, from the middle panels we
find that the $\Delta\Omega$ from B-DECIGO is actually 1 to 4 orders of
magnitude better than ET, especially in categories (b) and (c). \add{But in the most optimistic case where ET and 2 CEs  are operating together at their best designed sensitivity, the localization becomes comparable between B-DECIGO and 3G detectors.}  The merger time
uncertainty $\Delta t_c$ is larger for B-DECIGO than 3G detectors. However, B-DECIGO's
sub-second errors on $t_c$ are already sufficient for the preparation of EM
observations, not to mention the long enough early warning time from B-DECIGO
that the ground GW detectors cannot provide.

Note that the above analyses only compare the results from B-DECIGO to the 3G detectors. Other decihertz detectors such as DO-Optimal and DECIGO could detect
sources further than 3G detectors (see Table~\ref{tab:Number/yr}). \add{Meanwhile, for the B-DECIGO sources, DO-Optimal and DECIGO could yield higher SNR and more accurate localization.} 
More details are given in the next section. The general trend is that space-based decihertz detectors have
better localization ability and worse time of merger accuracy. 

\begin{figure}
\gridline{\fig{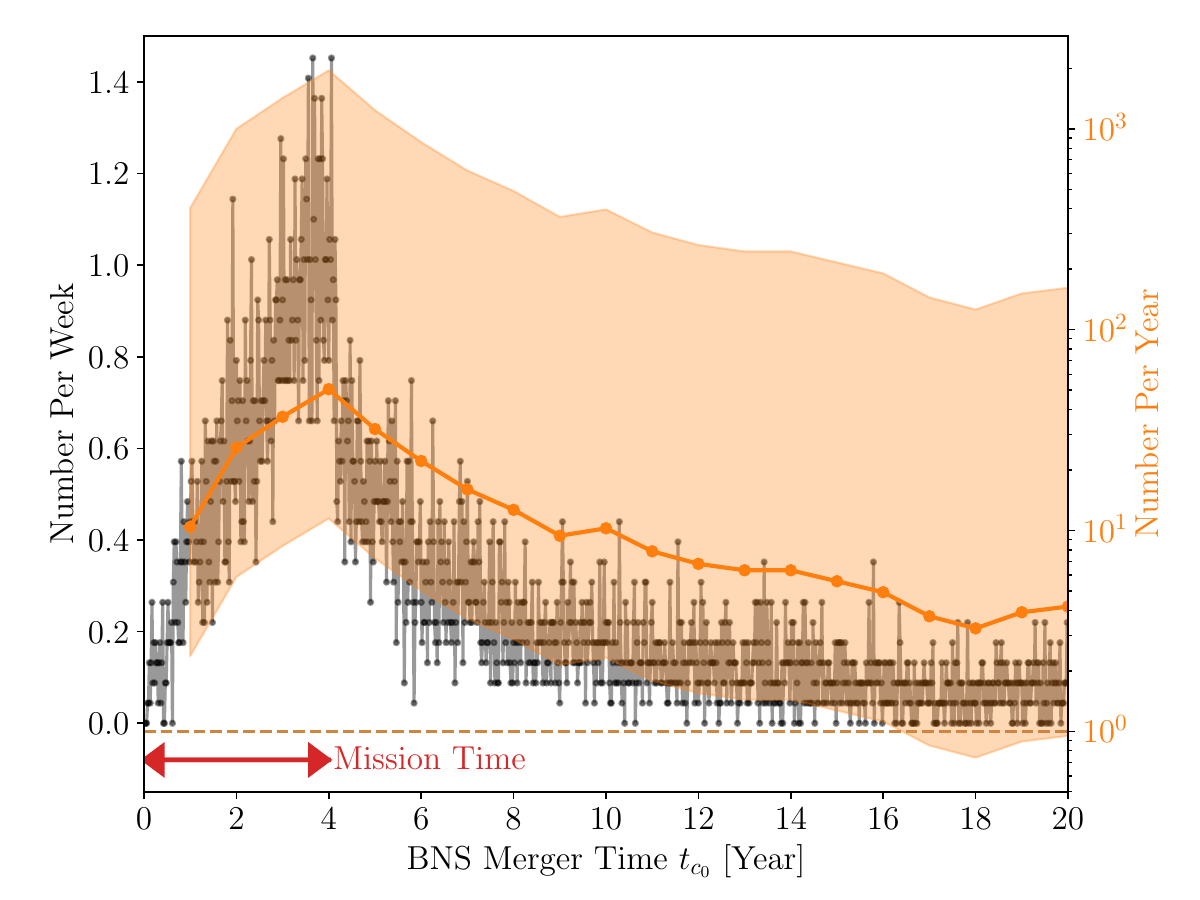}{0.45\textwidth}{}}
\vspace{-1.6cm}
\gridline{\fig{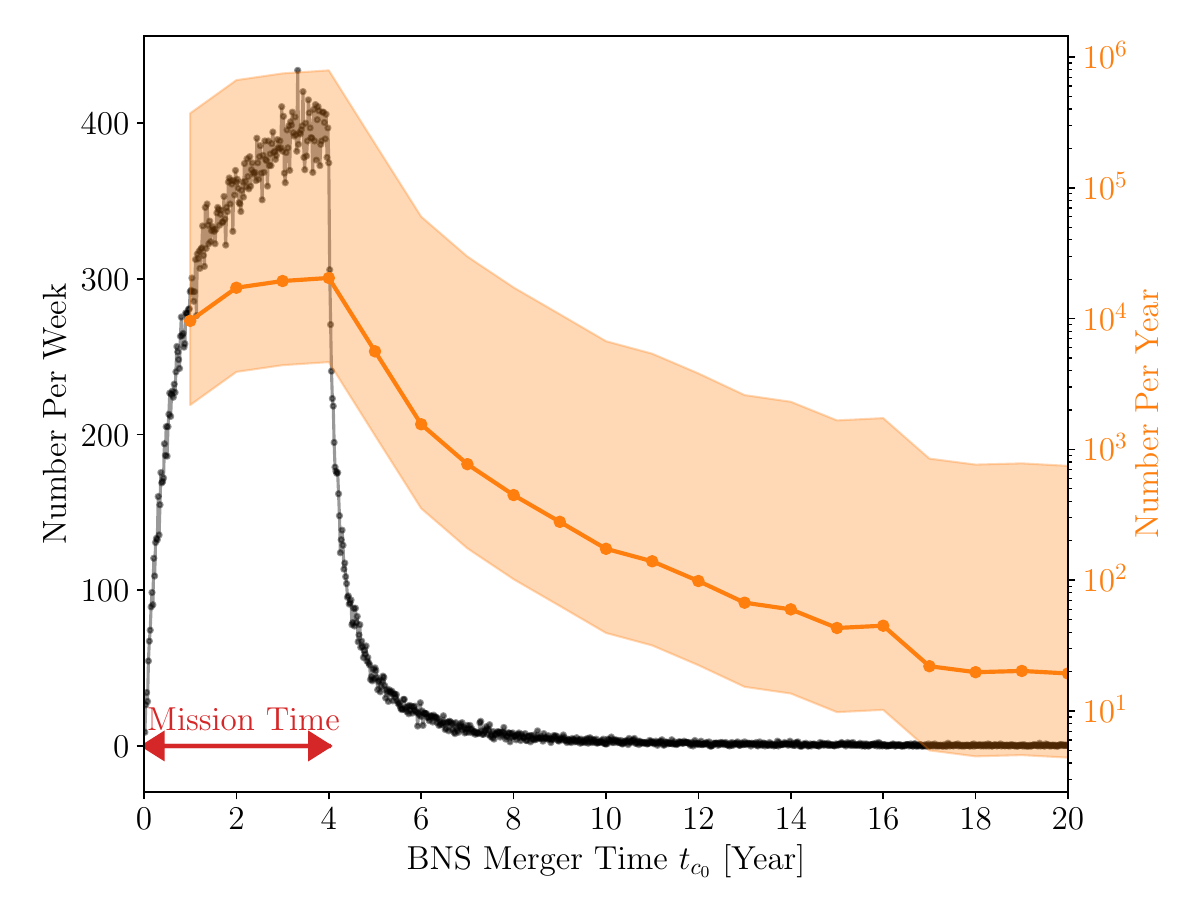}{0.45\textwidth}{}}
\vspace{-1.6cm}
\gridline{\fig{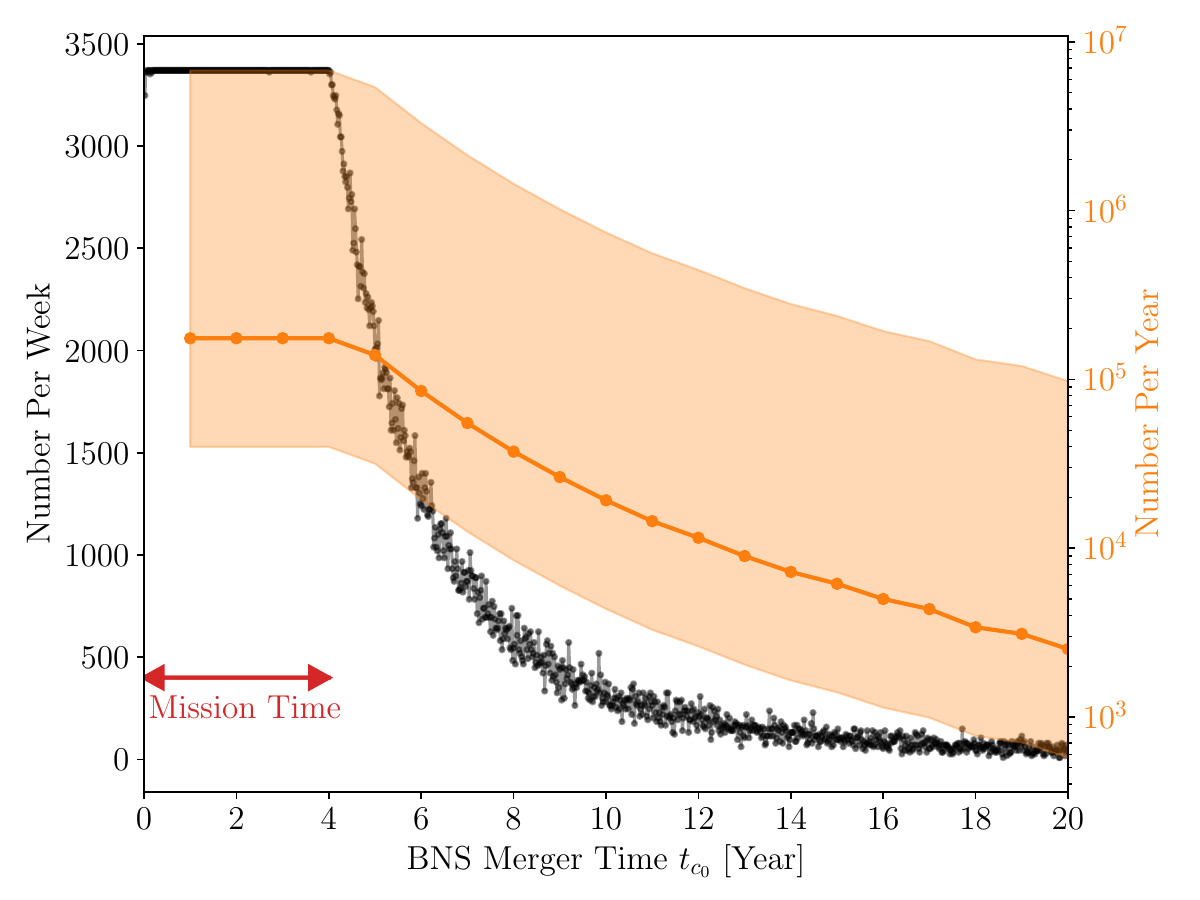}{0.45\textwidth}{}}
\vspace{-0.6cm}
	\caption{Same as Fig.~\ref{fig:BDEC_number}, for DO-Conservative ({\it
	top}), DO-Optimal ({\it middle}), and DECIGO ({\it bottom}).}
\label{fig:DO:CON:OPT:DECIGO}
\end{figure}
\begin{figure}
	\includegraphics[width=8.5cm]{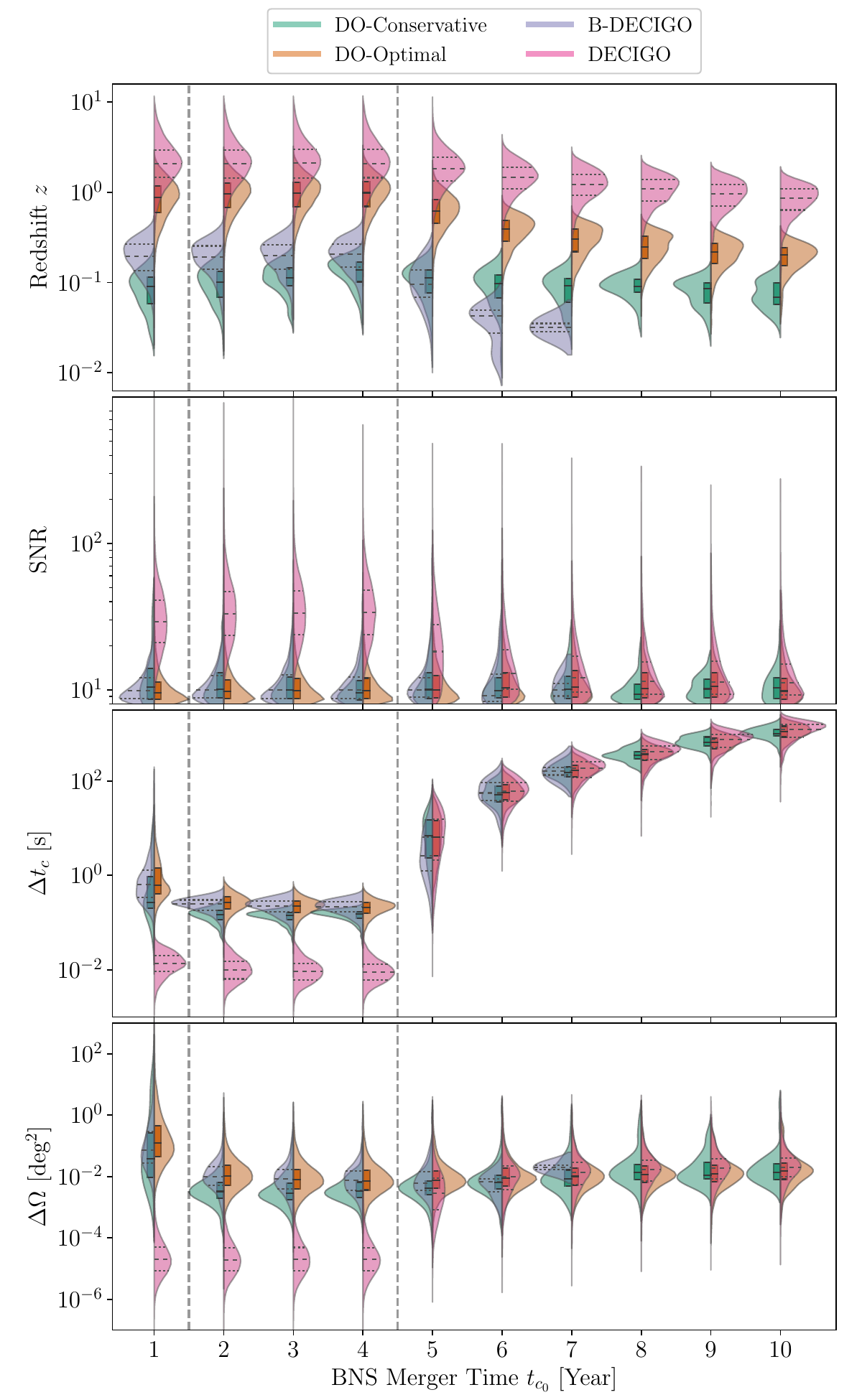}
	\caption{Probability distributions of redshift ({\it top}), SNR ({\it second panel}), $\Delta t_c$ ({\it third panel}), and $\Delta \Omega$ ({\it bottom}) of the
	detected sources in DO-Conservative ({\it green}), DO-Optimal ({\it orange}), B-DECIGO
	({\it blue}), and DECIGO ({\it pink}) as a function of BNS merger time $t_{c_0}$. The horizontal dashed lines and the box plots on each violin plot show the quartiles of each distribution.
	The original sources are generated according to the population model ``SFR14'' and the two vertical dashed
	lines divide BNSs into categories (a), (b), and (c).}
    \label{fig:compare_dets}
\end{figure}

\begin{figure*}
	\includegraphics[width=7.8cm]{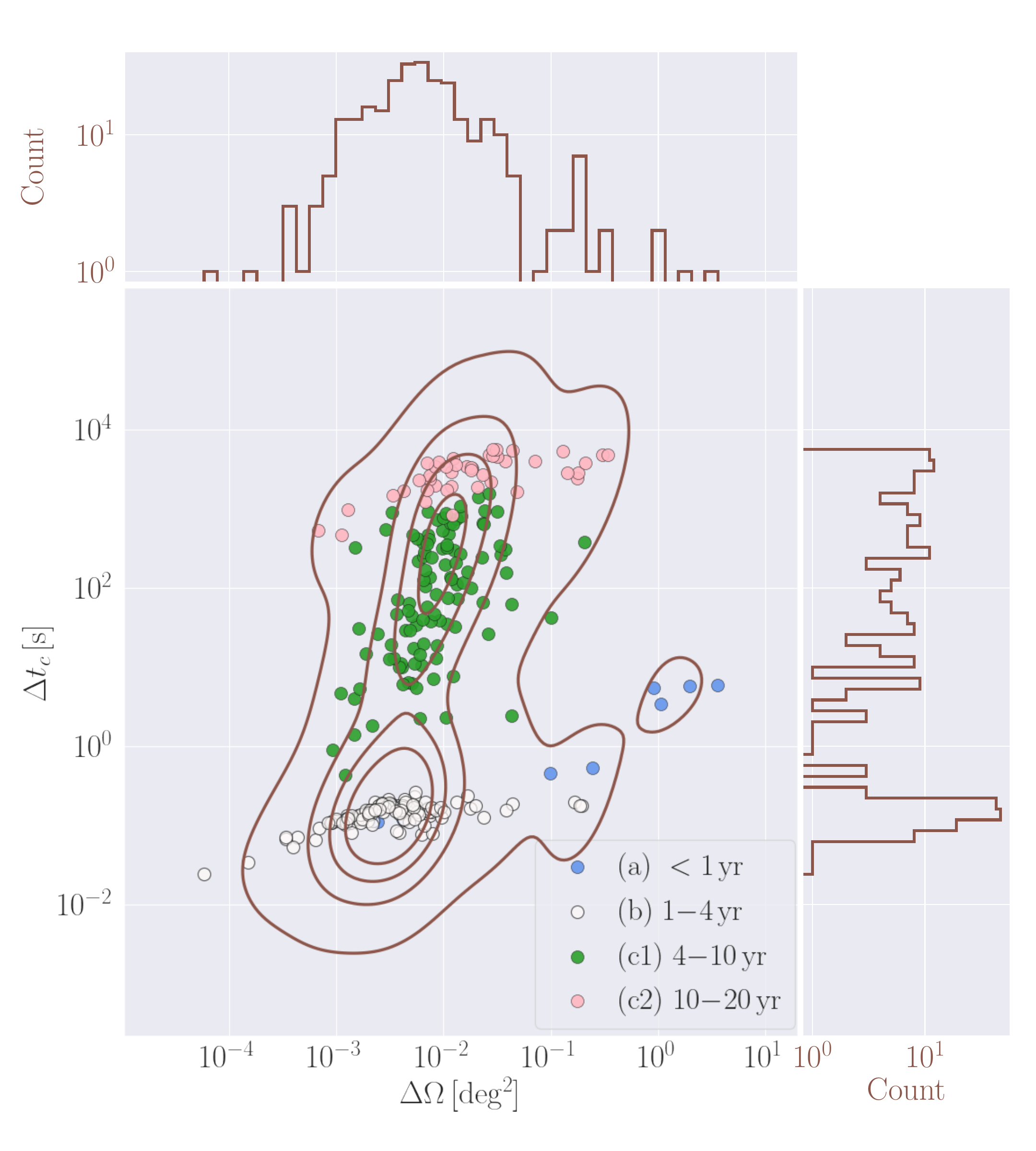} \hspace{0.7cm}
	\includegraphics[width=7.8cm]{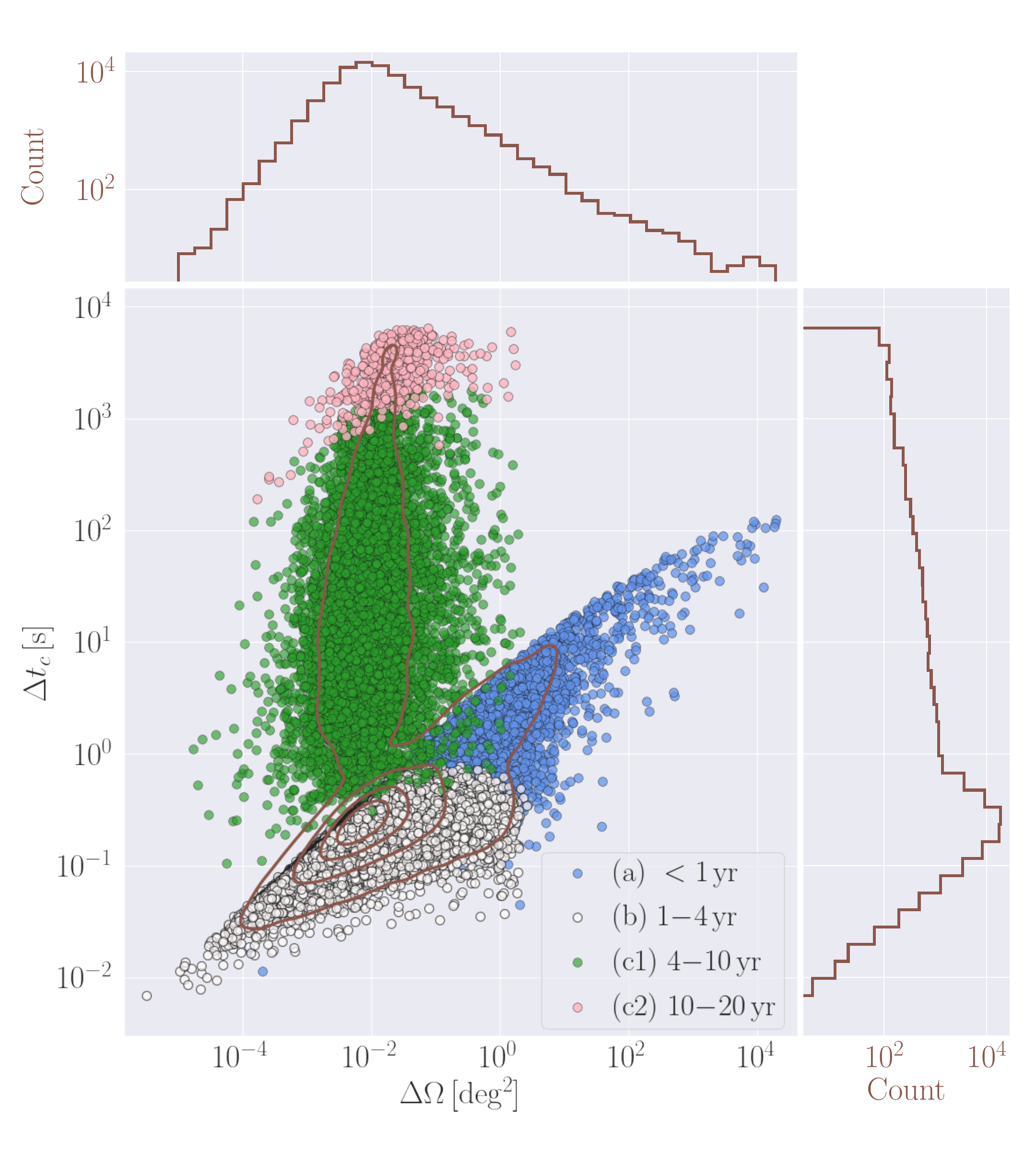} \\
	\includegraphics[width=8.5cm]{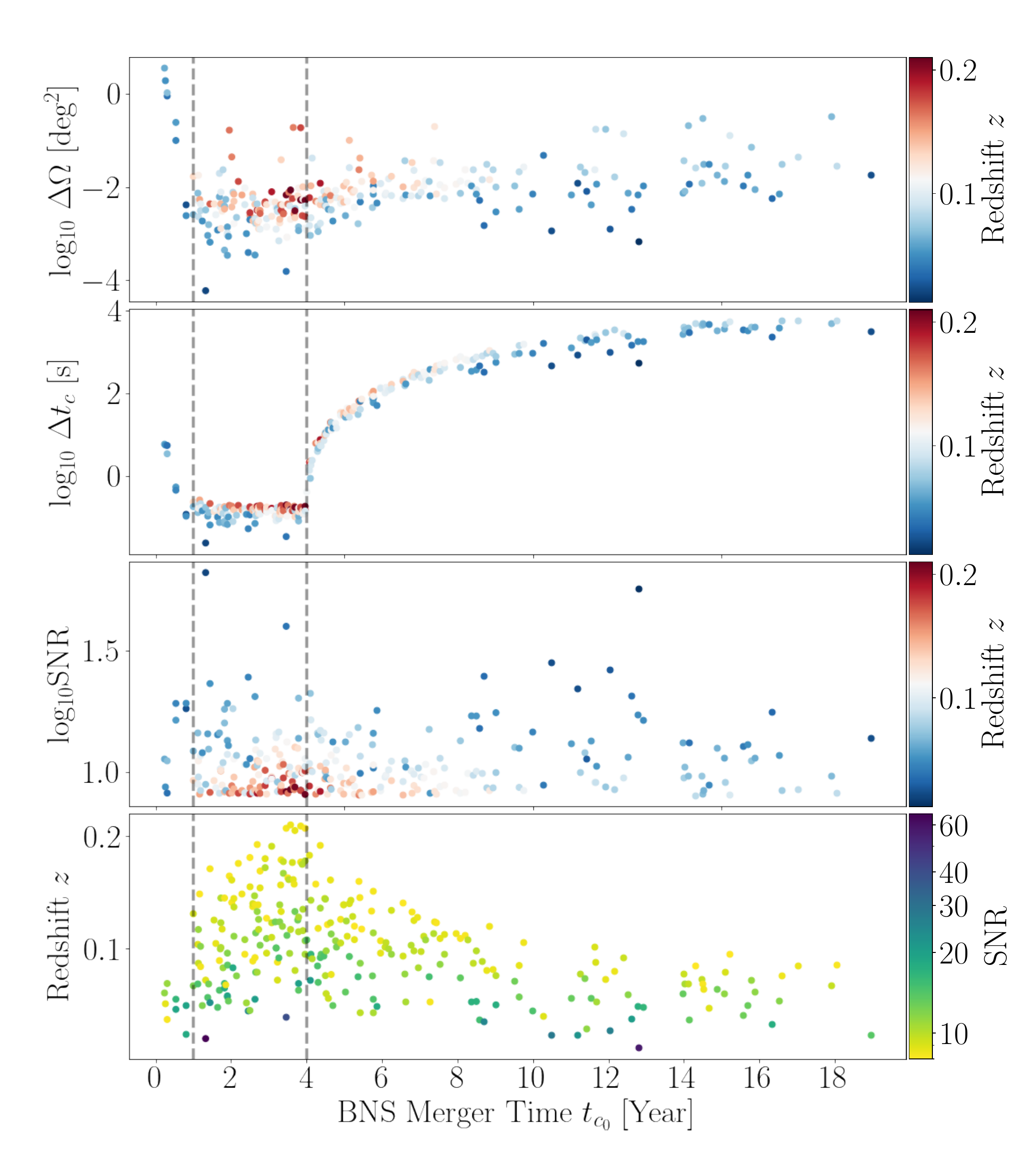}
	\includegraphics[width=8.5cm]{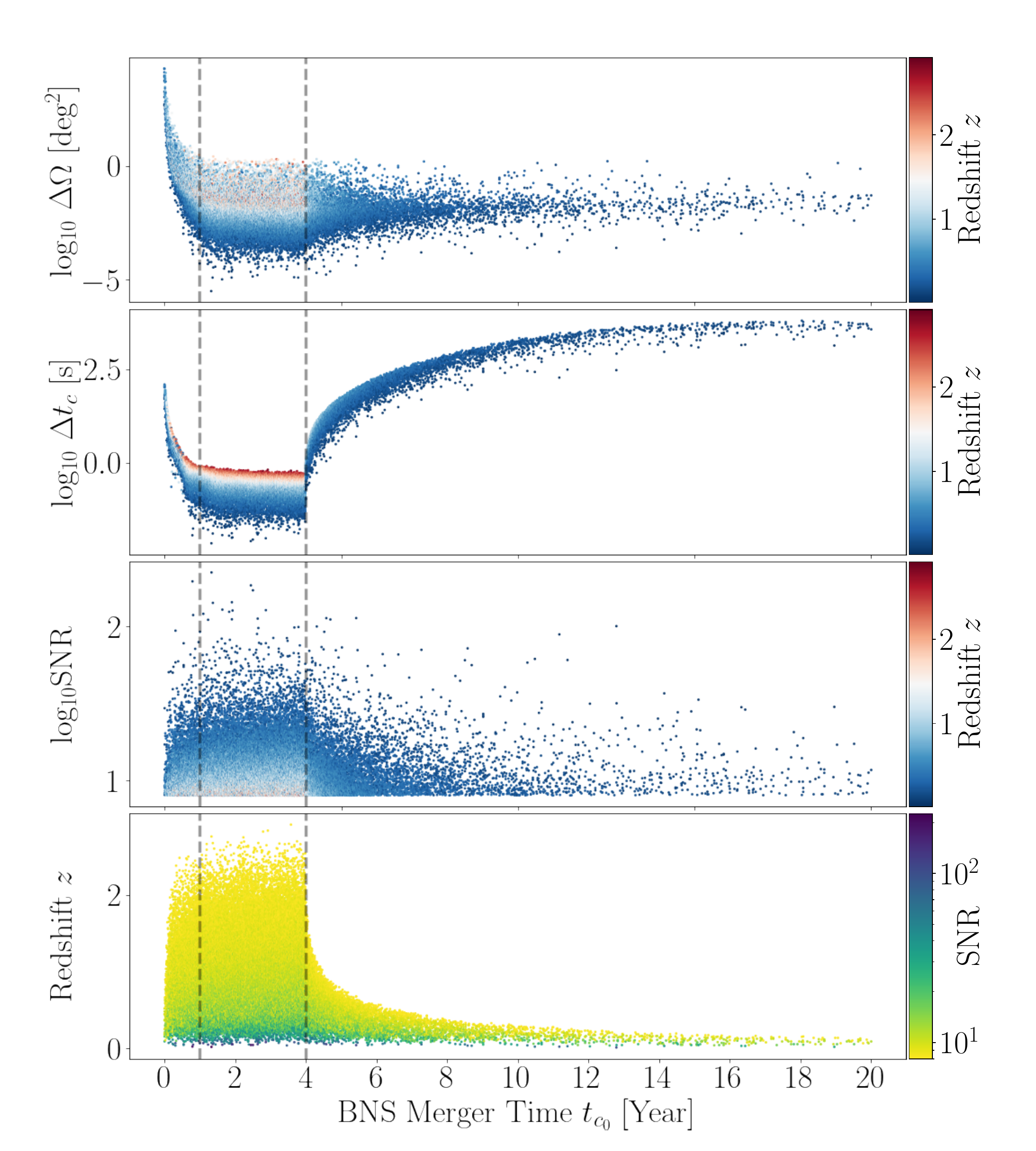}
    \caption{({\it Upper}) same as Fig.~\ref{fig:BDEC_dist}, for DO-Conservative ({\it left}) and DO-Optimal ({\it right}). 
	({\it Lower}) same as Fig.~\ref{fig:BDEC_4panel}, for DO-Conservative ({\it left}) and DO-Optimal ({\it right}). }
	\label{fig:DO:combined:new}
\end{figure*}


\section{Other decihertz detectors}
\label{sec:dets}

We begin by analyzing the common features.  Figure~\ref{fig:DO:CON:OPT:DECIGO}
shows the yearly numbers of detections from DO-Conservative, DO-Optimal, and
DECIGO. A summary is given in Table~\ref{tab:Number/yr}. We notice that all
these three detectors could observe sources with $t_{c_0}$ up to $20\,$yr,
since they all have a lower noise level at the frequency range of BNS inspirals for sources in
category (c). The numbers of detections, however, vary significantly due to
different levels of the total noise.

Figure~\ref{fig:compare_dets} shows the probability distributions of the redshift~$z$, SNR, 
$\Delta t_c$, and $\Delta \Omega$ from the 4 decihertz detectors. The top panel
shows the distributions of the redshift and we also record the largest redshift
$z_{\rm max}$ in Table~\ref{tab:Number/yr}.  Instead of using the sky-averaged
estimations, our recorded $z_{\rm max}$ takes into consideration of the
directional dependency of the source, therefore it represents the farthest horizon
distance. 
From the last two panels in Fig.~\ref{fig:compare_dets} we find that the
categories (a), (b), and (c) are more dominant than the
detector sensitivity in determining the distributions of $\Delta t_c$ and
$\Delta\Omega$. In categories (b) and (c), $\Delta\Omega$ of the sources
is always less than $\mathcal{O}(1)\,\rm{deg}^2$, regardless of the qualities of the
detectors, and the timing accuracies of most sources in categories (c) are
$\Delta t_c>1\,\rm s$.

We now
explore the specialties of decihertz GW detectors one by one.
Compared to B-DECIGO, DO-Conservative has a higher noise level at $f>$ 0.2 Hz,
leading to a significant drop in the number of detections. The weekly detection
rates hardly reach one. However, DO-Conservative has a better sensitivity at
lower frequencies, which greatly boosts the number of detections in category (c). Thus, it can observe sources that will
merge more than dozens of years after the mission ends. Exclusive
characteristics for DO-Conservative could be found in the left panels of 
Fig.~\ref{fig:DO:combined:new}.

Using DO-Optimal, the weekly detection rates of the sources with
$t_{c_0}<4\,{\rm yr}$ gradually increase, from 0 to 400, but then drop
drastically in category (c). Approximately 2
years after the detector closes down, the weekly number of detections falls to be fewer than 10. Though DO-Optimal is an upgraded version of DO-Conservative, the parameter distributions for both of them are similar, \add{due to its further horizon}. Exclusive characteristics for DO-Optimal could be found in the right panels of 
Fig.~\ref{fig:DO:combined:new}.

DECIGO is the most powerful detector in the decihertz band that we consider. Except for the first
$\sim 10$ weeks, 
DECIGO will detect all the sources with $t_{c_0}<4$ years. Meanwhile, because of its full design
with a very high sensitivity, both categories (a) and (b) have the same
level of localization and timing abilities, which are orders of magnitude more
accurate than the other three detectors. 

\section{\add{Distinguishing} population models}
\label{sec:pops}

In Fig.~\ref{fig:pops} we have presented the redshift distributions of the 4
BNS population models, as well as the distributions of the merger events detected by B-DECIGO and DO-Optimal in their last year of operation ($3\,\rm{yr}<t_{c_0}\leq4\,$yr)  for each
population model \add{and discuss how detectors could help distinguish these models}.

In general, though different populations could be clearly distinguished by
their original redshift distributions, it is hard for us to tell the differences \add{from} the detectable sources---they are unlike the original
distributions due to the horizon distance of B-DECIGO and DO-Optimal.  Though a
dedicated Bayesian analysis might be able to give confident inference to them. 
Model ``Oce.High'' has the largest number of sources, leading to the largest detection number for both detectors, which makes it easily recognized. 
Models ``SFR14'', ``LN'', and ``Stan.High'', however, could not be
distinguished by B-DECIGO. The horizon distance of B-DECIGO is $z_{\rm max} \sim
0.45$ while sources from these three models have similar distributions before $z
= 1$. However, after $z=1$, the number of mergers for model ``SFR14'' raises, while for models ``LN'' and ``Stan.High'', it drops sharply and remains stable, respectively.
\add{Since the horizon of DO-Optimal reaches $z_{\rm
max}\sim3$, it helps distinguish the other 3 models.
Sources detected from model ``SFR14''
outnumber the other two, whereas model ``LN'' has several orders of
magnitude fewer detections than model ``Stan.High'' in the last 4 redshift bins in Fig.~\ref{fig:pops}.}
\add{Though it is hard to determine the formation channel only by the detection numbers, detectors with larger horizons (such as DO-Optimal/DECIGO) have the ability to discriminate the redshift distribution, e.g.\ between ``SFR14'' and ``LN''. By identifying the peak of the merger rates, space-based decihertz detectors could give hints on the time delay between star formation and binary mergers.


As we will see in the next section, population models have different contributions to the confusion noise and they will affect the detection rates of various detectors differently. The influence of  confusion noise could be another way to discriminate population models.}

\begin{table}
	\def\setstretch{1.25}
	\centering
	\caption{The yearly number of detections of the sources with $t_{c_0}<1\,$yr
	in category (a), with (numerator) or without (denominator) the consideration
	of the confusion noise from GW foreground. We show their ratios in brackets. }
	\label{tab:Conf}
	\setlength{\tabcolsep}{0.96mm}{\begin{tabular}{lcccc} 
		\hline
		\hline
		    & B-DECIGO & DO-Conservative & DO-Optimal   \\
		  		 \hline
SFR14 & 124/127 (98\%) & 8/8 (100\%)& 5600/9600 (58\%)   \\
LN & 159/160 (99\%) & 13/13 (100\%) & 6700/8700    (77\%)\\
Stan.High & 51/124 (41\%) & 4/9 (44\%)& 211/6300 (3.3\%)    \\
Oce.High & 48/349 (14\%) &  6/31 (19\%)& 91/22000 (0.7\%) &   \\
		\hline
	\end{tabular}}
\end{table}

\section{The impact of confusion noise}
\label{sec:conf}

All the results above leave out the consideration of the confusion noise
\citep{Christensen:2018iqi,Barish:2020vmy}. In reality, the stochastic GWs from
the undetectable sources could leave imprints in the detectors, which are determined by the fractional energy density $\Omega_{\rm GW}$. The amplitude
of $\Omega_{\rm GW}$ contributed from double compact objects (DCO) is about
$10^{-12}$--$10^{-9}$ at 1\,Hz depending on the population properties. It
transforms to the confusion noise shown in the grey dotted lines in
Fig.~\ref{fig:curve}. Appendix~\ref{sec:conf_eq} shows the details of calculation.

It has been shown that, the confusion noise from DCOs could be subtracted out by an
iteration scheme and a global fit \citep{Cutler:2005qq}. However it depends severely
on the specific population model, the sensitivity of the detector and its operation
time, which leads to large uncertainties in the calculation. Nevertheless, as shown by
\citet{Kudoh:2005as} and \citet{Yagi:2011wg}, the full design of DECIGO is adequate for
subtracting out the binary foreground. 

Here we take a conservative approach.  We have calculated the confusion noise
based on the population models implemented in our paper (see
Table~\ref{tab:Omega}),
 and estimated the number of detections with or without
the confusion noise in Table~\ref{tab:Conf}. 
Note that we do not include DECIGO
since it can identify all the sources.  We do not have estimations on
$\Omega_{\rm GW}^{\rm NSBH}$ and $\Omega_{\rm GW}^{\rm BBH}$ for population
models ``SFR14'' and ``LN'', since they are phenomenological models fitted to
the observations. But for models ``Stan.High'' and ``Oce.High'', we obtain an
estimation on $\Omega_{\rm GW}^{\rm NSBH}$ and $\Omega_{\rm GW}^{\rm BBH}$ by
their NSBH/BBH population models from \citet{Dominik:2013tma}. Such a difference
leads to a significant distinction in detection rates.

Table~\ref{tab:Conf} presents the number of detections of the sources that will
merge in the first year with or without the confusion noise. Considering the
subtraction scheme, the actual numbers of detections should be somewhere in
between.  We notice that the confusion noise purely from BNS
populations only slightly affects our results, especially for B-DECIGO and
DO-Conservative where the effect is smaller than $2\%$. However, since the BBH foregrounds are orders of magnitude larger than that of BNS, they will affect
the numbers of detections severely. B-DECIGO and DO-Conservative will miss
$\sim 50\%$ of the detectable sources for the ``Stan.High'' model and $\sim
80\%$ for the ``Oce.High'' model.  The DCO foreground has the largest impact on
DO-Optimal. Its number of detections declines drastically no matter in only the BNS
foreground or the whole DCO foreground, though the absolute numbers are still
greater than those of B-DECIGO and DO-Conservative.

The above results are just for a conservative reference, since we
did not subtract the confusion noise due to the complexity of population models
and observation periods. For detectors with closer horizon distances, such
as DO-Conservative, few subtractions need to be done and our results are more close to reality.

\section{Discussions}
\label{sec:sum}
We provide the detections and early warning predictions of different BNS
populations to be observed by \add{space-based} decihertz GW observatories with realistic simulations.  We show that the detected sources could be divided into 3 categories based on their properties on a
$\Delta\Omega$-$\Delta t_c$ map, which is  determined by the specifics of
space-based heliocentric-orbit detectors.  Sources in category (a) that merge quickly have less
accurate localization and timing precision. Sources in category (b) that merge
within 1--4 years after the launch have stable and the best parameter estimation
results. Sources in category (c) that do not merge within the mission time could
still offer early warning alerts.

\add{We also discuss the landscapes of EM follow-up observations for the three categories and compare the Monte Carlo simulation results from B-DECIGO with ET, ``ET+CE'',}
DO-Conservative, DO-Optimal, and DECIGO. Furthermore, we provide the detection prospects for 4 different
population models and discuss the influence of the confusion noise.

With such strong localization capability from decihertz space-based GW detectors, joint
detections with EM telescopes and satellites become possible. Synergy observations might be carried out with, e.g., SWIFT
\citep{SwiftScience:2004ykd}, GECAM \citep{Zhang:2018hzq}, eXTP
\citep{eXTP:2018kws}, EP \citep{YuanEP}, THESEUS \citep{Ciolfi:2021gzg}, WFST
\citep{shi2018}, Mephisto \citep{lei2021}, ZTF \citep{Graham:2019qsw}, and JWST
\citep{Gardner:2006ky}. For the EM facilities that have field-of-views larger
than squares-of-degree level, joint detections will almost always succeed as
long as EM counterparts reach above their detection thresholds. Therefore,
future multi-messenger astronomy using decihertz detectors with EM follow-ups
will be very promising, and provides interesting science outcomes.

\section*{Acknowledgements}

\add{We thank the anonymous referee for helpful comments.}
This work was supported by the National Natural Science Foundation of China
(11975027, 11991053, 11721303), the National SKA Program of China
(2020SKA0120300), 
the Max Planck Partner Group Program funded by the Max Planck Society, 
and the High-Performance Computing Platform of Peking University.  
Y.K. acknowledges the Hui-Chun Chin and Tsung-Dao Lee Chinese Undergraduate 
Research Endowment (Chun-Tsung Endowment) at Peking University.
\facilities{\add{DECIGO, B-DECIGO, DO, ET}}

\appendix
\section{Merger rate calculation}
\label{sec:rate_eq}

In models ``SFR14'' and ``LN'', we assume  that the BNS merger rate density
evolves with redshift $z$ via
\begin{equation}
	\mathcal R(z) = \mathcal R_0 f(z)\,,
	\label{eq:rate}
\end{equation}
where $\mathcal R_0$ is the local merger rate and $f(z)$ is the corresponding
normalized (dimensionless) redshift distribution model. For models ``Stan.High''
and ``Oce.High'', we use the $\mathcal R(z)$ provided by \citet{Dominik:2013tma}
directly. The number of merger events up to redshift $z_{\rm lim}$ within a
period $T$ can then be obtained with 
\begin{equation}
	N(z_{\rm lim})=T \times \int_{0}^{z_{\rm lim}} \frac{4 \pi
	D_{L}^{2}(z) \mathcal{R}(z)}{(1+z)^3 H(z)} \mathrm{d} z\,,
	\label{eq:num}
\end{equation}
where $T=20\,\rm yr$ (i.e., $\sim 20\times52\,\rm{weeks}$),  $z_{\rm
lim} = 10$  by our choice,  and $H(z) \equiv H_{0}
\sqrt{\Omega_{\mathrm{M}}(1+z)^{3} + \Omega_{\Lambda}}$ is the Hubble parameter
at redshift $z$. 

\section{GW foreground from unresolved DCO systems}
\label{sec:conf_eq}

\begin{table}
   \def\setstretch{1.25}
	\centering
	\caption{The values of $\Omega_0^{\rm DCO}$ (in the unit of $10^{-12}$) in
	Eq.~\eqref{eq:Omega} for different population models. The  population models
	``Stan.High'' and ``Oce.High'' for NSBHs and BBHs are  taken from
	\citet{Dominik:2013tma}.}
	\label{tab:Omega}
	\begin{tabular}{lccc} 
		\hline
		\hline
		    & BNS & NSBH & BBH   \\
		 \hline
SFR14 & 4.15 & -- & --    \\
LN & 2.66 & -- & --    \\
Stan.High & 2.40 & 4.50 & 405   \\
Oce.High & 11.0 & 15.8 & 2227   \\
		\hline
	\end{tabular}
\end{table}

At the decihertz band, a stochastic GW from DCOs is dominated by their inspiral
stages, and characterized by its fractional energy density $\Omega_{\rm
GW}$ per logarithmic frequency interval,
\begin{equation}
	\Omega_{\mathrm{GW}}(f) \equiv \frac{1}{\rho_{\mathrm{c}}} \frac{\mathrm{d}
	\rho_{\mathrm{GW}}(f)}{\mathrm{d}\ln f}\,,
\end{equation}
where $\rho_{c}=3 H_{0}^{2} /(8 \pi)$ is the critical energy density of the
Universe. The GW foreground by astrophysical compact binaries is given by
\citep{Phinney:2001di}
\begin{equation}
	\Omega_{\mathrm{GW}}^{\mathrm{DCO}}(f)=\frac{8 \pi^{5 / 3}}{9}
	\frac{1}{H_{0}^{2}} \mathcal{M}^{5 / 3} f^{2 / 3} \int_{0}^{\infty} {\rm d}
	z \frac{{\mathcal R}(z)}{(1+z)^{4 / 3} H(z)}\,.  \label{eq:Omega}
\end{equation}
Taking a specific population model and integrating Eq.~\eqref{eq:Omega}, one
derives,
\begin{equation}
\Omega_{\rm G W}^{\rm D C O}(f)= \Omega_{0}^{\rm D C
O}\left(\frac{H_{0}}{67.4\,\rm km\,s^{-1}\,Mpc^{-1}}\right)^{-3} 
\left(\frac{\mathcal{M}}{\mathcal{M}_{\rm D C O}}\right)^{5 /
3}\left(\frac{f}{1\, {\rm Hz}}\right)^{2 / 3}\,,
\end{equation}
where $\mathcal{M}_{\rm D C O}$ is the chirp mass of the sources in the
population. We choose $\mathcal{M}_{\rm BNS} = 1.22 \, M_{\odot} $,
$\mathcal{M}_{\rm NSBH} = 6.09\, M_{\odot} $, and $\mathcal{M}_{\rm BBH}= 24.5\,
M_{\odot}$ in the calculation. The $\Omega_0^{\rm DCO}$ for various population
models are listed in Table~\ref{tab:Omega}. 

The total GW foreground spectrum $S_h$ is then
\begin{equation}
	S_{h}^{\rm DCO}(f)=\frac{4}{\pi} f^{-3} \rho_{c} \Omega_{\mathrm{GW}}^{\rm
	DCO}(f)\,,
\end{equation}
which is plotted as dotted grey lines in Fig.~\ref{fig:curve} for model
``Stan.High''. The confusion noise of unresolved systems will then modify the
noise spectrum of the detector via
\begin{equation}
	S_{\rm n}^{\rm total}(f) = S_{ h}^{\rm DCO}(f) + S_{\rm n}^{\rm
	instrument}(f) \,.
	\label{eq:Sn}
\end{equation}
We use $S_{\rm n}^{\rm total}$ to explore the effects of the confusion noise in
Sec.~\ref{sec:conf}.

\bibliographystyle{aasjournal}
\bibliography{refs} 

\end{document}